\documentclass[12pt]{article}
\usepackage{epsfig}
\newcommand{\as}{\alpha_s} 
\newcommand{\gsim}{\stackeven{>}{\sim}}
\newcommand{\stackeven}[2]{{{}_{\displaystyle{#1}}\atop\displaystyle{#2}}}

\begin{document}

\title {
\begin{flushright}
{\small INT--PUB--03--11}\\
\end{flushright}
{ Particle Correlations at High Partonic Density }}
\author{
{Kirill Tuchin} \\[5mm]
{\it\small Institute for Nuclear Theory, University of Washington,
Box 351550}\\
{\it\small Seattle, WA 98195, USA}
}

\date{}

\maketitle

\begin{abstract}
 We discuss manifestations of the particle correlations at high
partonic density in the heavy-ion collisions at RHIC. In particular, we
argue that the elliptic flow variable $v_2$ is dominated by particle
correlations at high $p_T$.
\end{abstract}

\maketitle

\begin{figure}[b]
\begin{center}
\epsfig{file=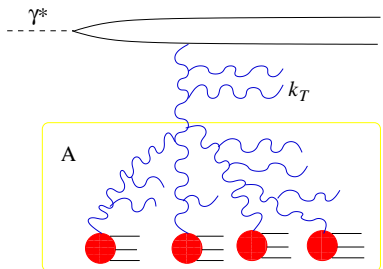, width=5cm}
\caption{Gluon production in DIS on a heavy nucleus.}\label{dis}
\end{center}
\end{figure}

Particle correlations at high partonic density (in Color Glass Condensate) 
are significantly different from those of the parton model. To illustrate 
this consider gluon production in Deep Inelastic Scattering on a 
heavy nucleus $A\sim 1/\as^6$ at high energies $x\sim e^{1/\as}$, 
Fig.~\ref{dis}. 
It was proved in \cite{DIS} that the collinear factorization breaks down for 
this process. Instead $\gamma^* A$ cross section can be written in 
the $k_T$-factorized form  \cite{GLR}. This allows to introduce function 
$\phi(x,q_\bot)$ which encodes quantum evolution and multiple gluon 
rescattering in nucleus. $\phi(x,q_\bot)$ is simply related to the 
forward scattering amplitude which satisfies the QCD evolution equation 
for high partonic densities \cite{BK}. Solution to that equation implies 
that the scale inherent to function  $\phi(x,q_\bot)$ is $q_\bot^2\sim 
Q_s^2=\Lambda^2_\mathrm{QCD}A^{1/3} e^{4\as y}$. 
At high energies $Q^2_s\gg \Lambda^2_\mathrm{QCD}$, therefore one cannot 
neglect the virtuality of the $t$-channel gluons compared to the
momentum of the produced hard particle $k_T\gg 1$~GeV.
As the result the transverse momentum conservation does not require
anymore 
that the momentum of the hard jet be compensated by equally large 
momentum of another jet moving in the opposite direction in the
transverse plane (in the center-of-mass frame).

\begin{figure}
\begin{center}
\begin{tabular}{ccc}
\epsfig{file=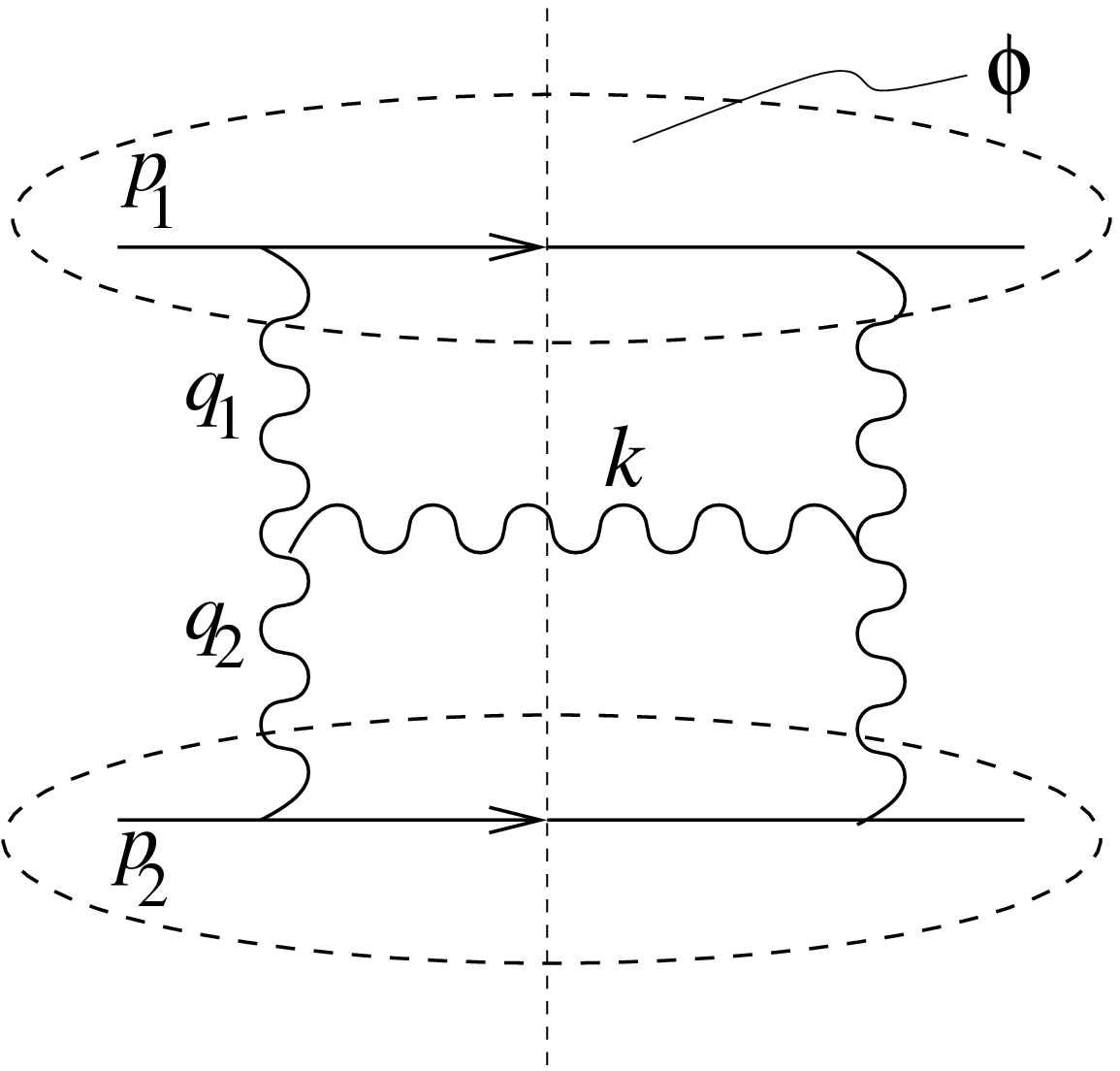, width=4cm}&\mbox{}&
\epsfig{file=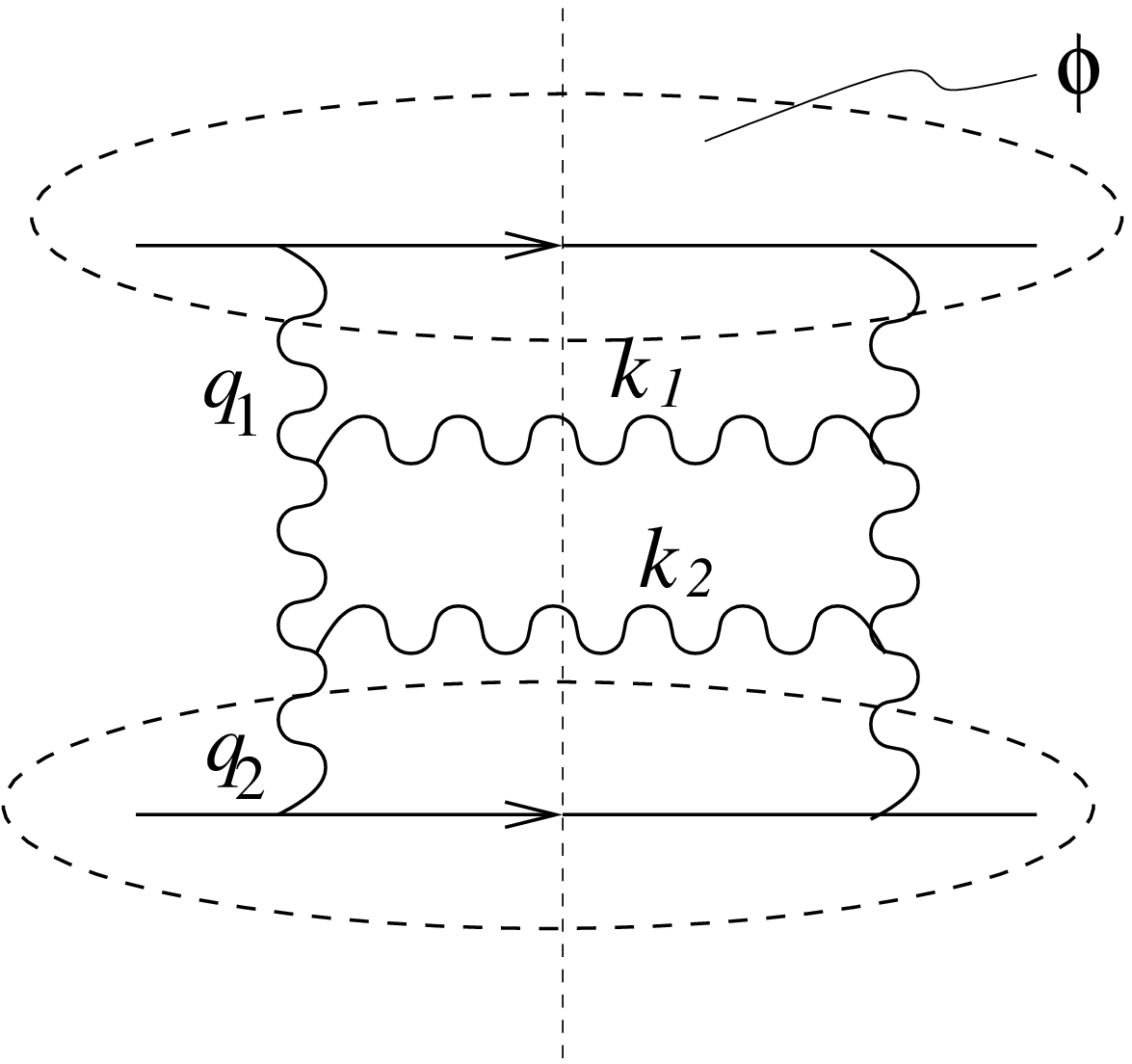, width=4cm}\\
a& &b
\end{tabular}
\caption{Inclusive (a) and double-inclusive (b) gluon production in
  AA}\label{prod}
\end{center}
\end{figure}

The qualitative picture of the gluon production in heavy-ion
collisions is pretty much the same as in DIS. Although the $k_T$
factorization has not been proved in this case, there are reasons to
believe that it is at least a fairly good approximation.
Feynman
diagram for the inclusive 
gluon production in AA collisions is shown in Fig.~\ref{prod}a.
To study the gluon correlations in heavy-ion collisions we
define the two-particle multiplicity distribution as
\begin{equation}\label{distr}
P(k_{1\bot},y_1; k_{2\bot},y_2)=\frac{dN}{d^2k_{1\bot}dy_1}
\frac{dN}{d^2k_{2\bot}dy_2} +
\frac{dN_\mathrm{corr}}{d^2k_{1\bot}dy_1 d^2k_{2\bot}dy_2},
\end{equation}
where the first term in the right-hand-side is just the square of the
diagram Fig.~\ref{prod}a which gives the uncorrelated piece, and the second
term is the diagram Fig.~\ref{prod}b which gives the correlated piece. 
The transverse momentum conservation applied to the later diagram
gives $|k_{1\bot}+k_{2\bot}|\simeq N Q_s$, where $N$ is the number of
gluons in the nuclei wave functions, cf.\ Fig.~\ref{dis}. Since $N$ can be
large, the collinear factorization can be violated at $k_\bot^2> Q_s^2$.
It was argued in \cite{geom} that the collinear factorization is recovered
when $k_\bot^2\gsim \tilde Q^2=Q_s^4/\Lambda_\mathrm{QCD}^2 A^{1/3}$. 
The preliminary dA data at $\sqrt{s}=200$ GeV suggest that $\tilde Q^2\gsim
Q_s^2$ at this energy. This means that at RHIC particles with momenta
$k_\bot^2\gg Q_s^2$ are correlated mostly back-to-back. It is important to
emphasize, that bulk of particles is produced with $k_\bot< Q_s$ 
and therefore, saturation plays crucial role in understanding of the total
multiplicity at RHIC.

\begin{figure}
\begin{center}
\begin{tabular}{ccc}
\epsfig{file=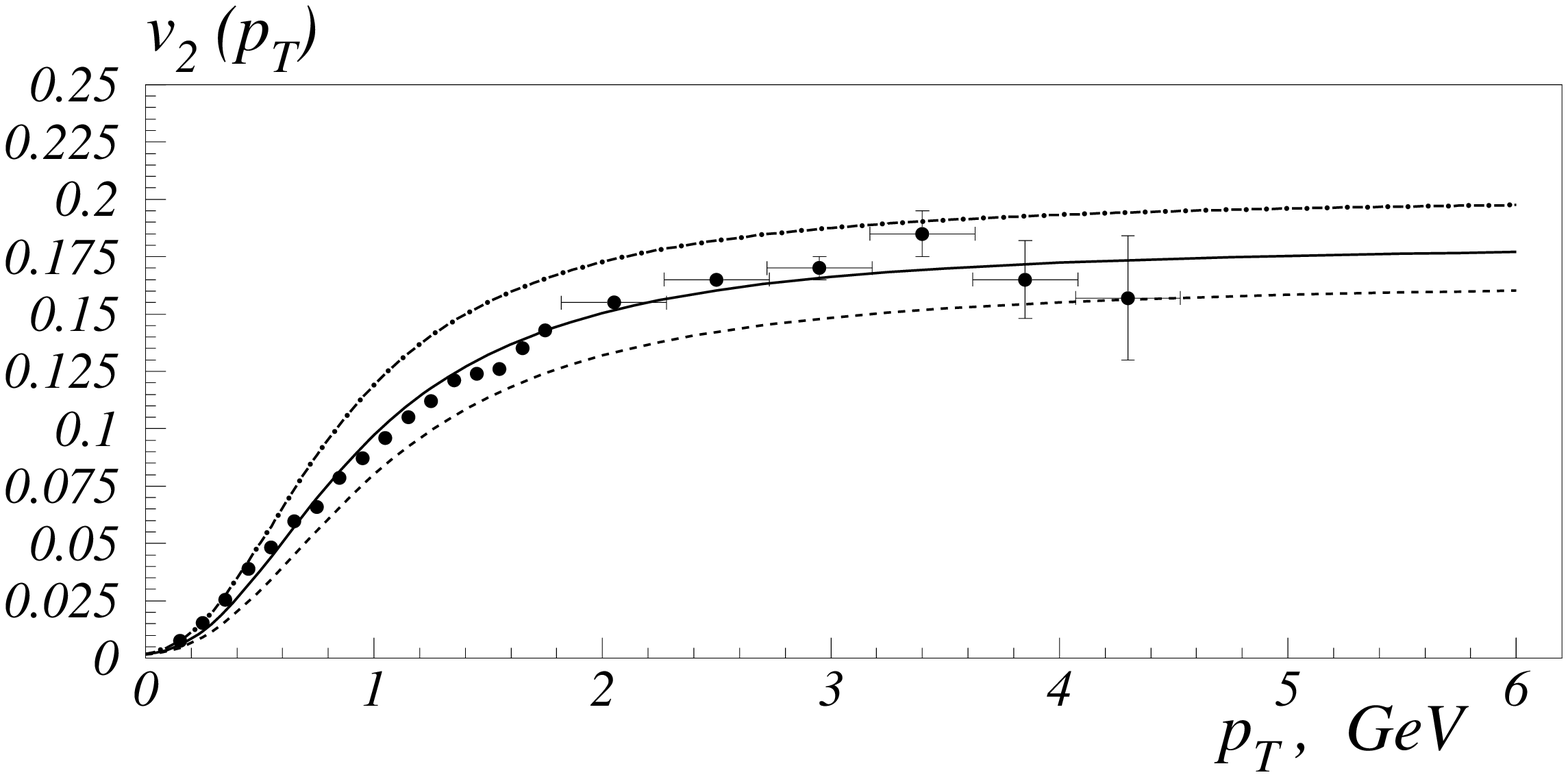,width=8cm,height=4cm}&
\epsfig{file=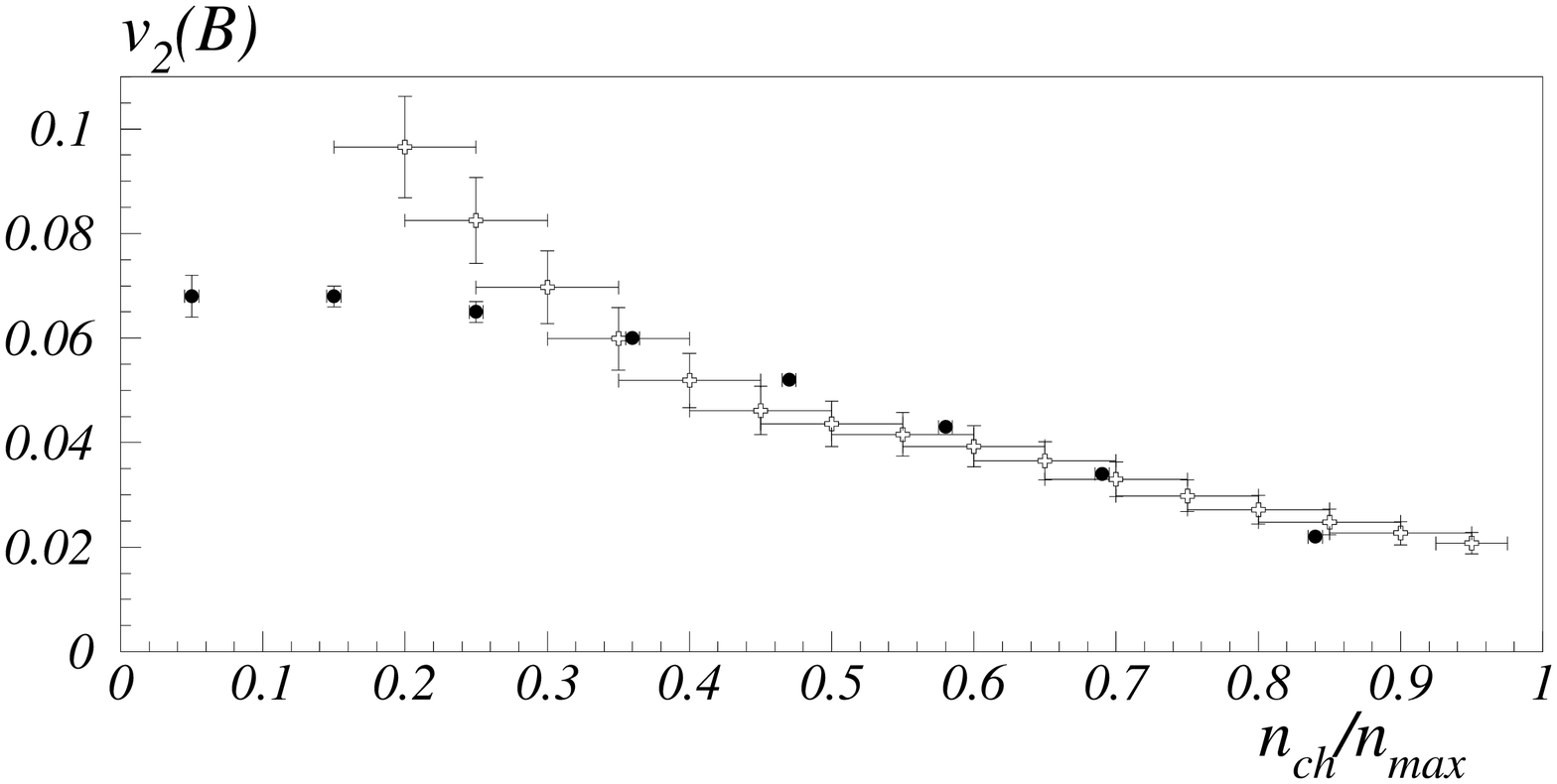, width=8cm,height=4cm}\\
a& b
\end{tabular}
\caption{Elliptic flow variable as a measure of the particle
  correlations in gold-gold collisions at $\sqrt{s}=130$~GeV vs STAR
  data.}\label{v2}
\end{center}
\end{figure}

Analysis of particle correlations with respect to the reaction plane
azimuthal angle defined as 
\begin{equation}\label{rp}
\tan 2\Psi_R=\frac{\sum_{i=1}^{'N}\sin2\phi_i}{\sum_{j=1}^{'N}\cos2\phi_i},
\end{equation}
shows that the angular momentum distribution of the large multiplicity
event $N\gg 1$ is given by \cite{cor}
\begin{equation}
\frac{dn}{d\phi_{p_T}d\Psi_R}=
\frac{1}{(2\pi)^2}[1+2 v_2(p_T,B)\Delta\cos2(\phi_{p_T}-\Psi_R)],
\end{equation}
where $\Delta$ is the reaction plane resolution. The 
$\cos2(\phi_{p_T}-\Psi_R)$ 
shape is the result of trivial autocorrelations
of each particle with itself. However the coefficient $v_2$ carries
information about the particle correlations.
The elliptic flow variable  $v_2$ is given by 
\begin{equation}
v_2(p_T)=\frac{\langle\cos2(\phi_1(p_T)-\phi_2)\rangle}
{\sqrt{\langle \cos2(\phi_1-\phi_2)\rangle}},
\end{equation}
where averaging over all events can be done using (\ref{distr}). 
The diagram Fig~\ref{prod}a entering the distribution function (\ref{distr}) 
can be calculated in a framework
of the McLerran-Venugopalan model \cite{MV}, i.e.\ treating  the
nuclear color field in the Weisz\"acker-Williams approximation
\cite{v2}. 
 The diagram Fig.~\ref{prod}b requires the $\as$ quantum correction. For simplicity 
we assume that nuclei have cylindrical shape. We also neglect all
finite state interactions. The result of calculation  \cite{v2} is shown in
Fig.~\ref{v2}.
It is in a reasonable agreement with data.

We conclude that the particle correlations are essential to understand
the behavior of the elliptic flow extracted from current flow analysis
methods. It seems to account for most of the flow at large $p_T$. 
Elliptic flow appears to be sensitive to the saturation physics of the
early stages of the collision.

\vskip0.3cm
The author would like to thank Yuri Kovchegov for continuing fruitful 
collaboration.
  This work was
sponsored in part by the U.S. Department of Energy under Grant
No. DE-FG03-00ER41132.


\vskip0.3cm


\begin{thebibliography}{99} 

\bibitem{DIS} Y.~V.~Kovchegov and K.~Tuchin,
Phys.\ Rev.\ D {\bf 65}, 074026 (2002)


\bibitem{GLR} 
L.~V.~Gribov, E.~M.~Levin and M.~G.~Ryskin,
Phys.\ Rept.\  {\bf 100}, 1 (1983).

\bibitem{BK}
I.~Balitsky,
Nucl.\ Phys.\ B {\bf 463}, 99 (1996);
Y.~V.~Kovchegov,
Phys.\ Rev.\ D {\bf 60}, 034008 (1999)


\bibitem{LT}
E.~Levin and K.~Tuchin,
Nucl.\ Phys.\ B {\bf 573}, 833 (2000)
Nucl.\ Phys.\ A {\bf 691}, 779 (2001)
Nucl.\ Phys.\ A {\bf 693}, 787 (2001)

\bibitem{geom}
E.~Iancu, K.~Itakura and L.~McLerran, 
Nucl.\ Phys.\ A {\bf 708}, 327 (2002);
D.~Kharzeev, E.~Levin and L.~McLerran,
Phys.\ Lett.\ B {\bf 561}, 93 (2003)

\bibitem{cor}
Y.~V.~Kovchegov and K.~L.~Tuchin,
Nucl.\ Phys.\ A {\bf 717}, 249 (2003)

\bibitem{MV}
L.~D.~McLerran and R.~Venugopalan,
Phys.\ Rev.\ D {\bf 50}, 2225 (1994)
Phys.\ Rev.\ D {\bf 49}, 2233 (1994)

\bibitem{v2} Y.~V.~Kovchegov and K.~L.~Tuchin,
Nucl.\ Phys.\ A {\bf 708}, 413 (2002)



\end{thebibliography}
\end{document}